# Trapping absorbing and non-absorbing aqueous particle using a universal 4-arm Laguerre-Gaussian mode light trap

*Krispin M. Dettlaff, James Wenger, Grégory David, and Ruth Signorell*

Chemistry and Applied Biosciences Department, ETH Zürich, Vladimir-Prelog-Weg 2, CH-8093 Zürich, Switzerland

**Abstract:** Absorbing aerosols, such as brown carbon (BrC) and absorbing secondary organic aerosols (SOA), has attracted broad interest due to their importance for climate and human health. The pronounced time-dependence of light absorption during aging renders the precise estimation of their impact on global warming difficult. Single particle studies of such aerosols would be very useful to better understand their aging in the atmosphere through processes such as photochemistry. However previously proposed optical traps cannot continuously trap particles whose absorption state changes from strongly absorbing to non-absorbing or vice versa. Some of the traps presented can isolate absorbing and non-absorbing particles, but require mechanical alignment of the trap depending on the strength of particle absorption. However, mechanical realignment is not compatible with continuous trapping and observation. Here, we introduce a flexible optical universal trap which does not require mechanical realignment. The versatility of the trap relies on four trapping beams - either vortex Laguerre-Gaussian (LG) or fundamental Gaussian beams - which are modulated with a spatial light modulator (SLM), The performance of the trap is demonstrated by trapping different types of absorbing and non-absorbing particles. We also show that the trap can be used to observe the photochemical reaction of aqueous droplets containing fulvic acid, a common component of BrC. Digital holography measurements demonstrate that the confinement of the particles in the trap can be controlled by changing the orbital angular momentum (OAM) of the LG beams. The study also shows that spectroscopy measurements, such as fluorescence and Raman scattering, are possible in all configurations of the proposed trap.

## 1. Introduction

Atmospheric aerosols are present everywhere on Earth, both indoor and outdoor. They influence the radiative forcing and hence on global warming [1, 2], and impact human health. Atmospheric aerosols are difficult to understand in detail because they have a wide range of physicochemical properties and are involved in and modified by numerous aging processes such as phase transitions, photochemistry, and reactions with gases and radicals. However, a better fundamental understanding is crucial to improve climate models, assess their health impact and optimize their use in life-science industries [2-12, x]. This concerns important atmospheric aerosols such as desert dust, black carbon, brown carbon (BrC) particles and secondary organic aerosols (SOA) [8, 13-17] (some SOA can be or become BrC upon extensive light absorption). Among them, in particular BrC and SOA have been widely studied [8, 11, 18] [19, 20]. Photochemistry and exposure to atmospheric gases modify the light absorption properties of SOA and BrC during formation and aging [8, 16, 17, 21], making it difficult to estimate their impact on the Earth radiative forcing [8, 13, 18]. It is crucial to study particles isolated in air, as their aging processes and kinetics can substantially different from those of deposited particles, or bulk solution [16, 17, 22-27].

Particle trapping methods are very useful in this context, enabling isolation of a single aerosol particle in air for many days under well controlled environmental conditions (surrounding gases, relative humidity, light exposure, etc.)[28, 29]. Numerous trapping methods have been developed to isolate single particles in the air, including acoustic levitation [30], electromagnetic balances [31-33], and optical traps [34-38]. Among the those, optical traps are the only ones that allow trapping of submicron-size particles [37-40], which is the most relevant size range of atmospheric aerosols. Several optical traps have been developed to trap absorbing or non-absorbing particles [34, 36, 41-44]. However, none of those allows dynamical tuning of the trap properties, which is prerequisite for observing the properties of particles with strongly changing absorption properties, hence preventing important investigations, including the formation and aging of single BrC aerosols. The difficulty in trapping particles with different absorption properties arises from the fact that the dominant trapping forces experienced by absorbing and non-absorbing particles, respectively, are not the same. Non-absorbing particles are immobilized by the usual scattering and gradient optical forces [45] (Fig. 1a), which can be realized by using a single or multiple focused Gaussian beams [38, 40, 46-48]. Absorbing particles also experience scattering and gradient forces but here the photophoretic force dominates and dictates the particle motion in the trap (PF)[42]. The PF arises from the momentum transfer between the particle and the molecules in the surroundings, which is governed by the temperature of the particle surface. Hence, temperature gradients across different surface regions of the particle create an imbalance of momentum transfer to the particle which creates the PF (Fig. 1b) [41, 42, 49]. The PF becomes dominant when the particle shows significant absorption of the trapping laser light to create a substantial temperature gradient. For the same imaginary part of the refractive index ($k$), the PF increases with increasing particle size. Because the PF drives the particle away from regions of high light intensity, hollow beams are necessary to photophoretically trap absorbing particles [36, 41-43]. Different hollow beam traps have been used to trap absorbing particles [36, 41-43, 50]. Furthermore, so-called universal optical traps which can trap non-absorbing and absorbing particles have been presented using one or two trapping beams [36, 42]. However, these universal traps cannot trap highly absorbing and weakly absorbing particles without realignment of the trapping beams. The reason is that weakly absorbing particles experience a so-called negative PF pushing the particle in the direction opposite of the light propagation direction while strongly absorbing particle experience a positive PF pushing the particle in the direction of the light propagation direction [36]. Ongoing trap realignment - which requires mechanical adjustments for the mentioned traps – is not feasible for those traps as it would not only take too much time but also result in particle loss during realignment. Hence those universal traps are not well suited to study particles with strongly changing light absorption properties (e.g. from absorbing to non-absorbing or vice-versa). To our knowledge, so far photophoretic traps have only been demonstrated to trap solid absorbing particles but not aqueous droplets. Trapping of solid particles is less challenging as they are non-volatile and more thermally-stable. Being able to trap aqueous absorbing droplets is important from an atmospheric perspective, e.g. because the formation and ageing of BrC often occur in liquid organic and aqueous aerosol droplets [15-17, 51-55].

Here we introduce a new universal optical trap using four trapping beams. It allows us to trap particles independently of their light absorption properties (i.e. for any values of $k$) and for particle radii between a few hundred nanometers and several micrometers. A spatial light modulator (SLM) is used for dynamic tuning of the trap properties without any requirement for mechanical realignment. This is achieved through continuous tuning between Gaussian beams to trap non-absorbing particles and hollow-core Laguerre-Gaussian beams (LG) to trap absorbing particles. Tuning between both types of laser beams using the SLM takes less than 20 milliseconds. In this article, the term LG beam always refers to vortex beams which contain orbital angular momentum ($|l| \neq 0$) and not to the fundamental Gaussian mode ($l=0$), which is referred as Gaussian beam. The four pairwise counter-propagating trapping beams enable trapping of both weakly and highly absorbing particles when vortex beams are used. Since non-absorbing particles (for which the PF is significant compared to the force due to the Brownian motion but still weaker than the gradient and scattering forces) can be trapped with either both Gaussian and LG beams, our trap also covers this case. The ability to tune the trap quickly and without any mechanical adjustment is key for particles with fast changing light absorption properties.

To demonstrate the performance of the new universal trap, we immobilized aqueous droplets containing fulvic acid, a typical BrC proxy [56], and observed their photochemical degradation. Fluorescence measurements revealed that the droplets remain stably trapped even when the absorbing fluvic acid was completely bleached, i.e. the droplet become non-absorbing. We quantify the confinement of absorbing and non-absorbing particles in the trap using in-line holographic imaging. To our knowledge, this is the first quantification of a universal trap confinement. We demonstrate how the confinement can be controlled by changing the orbital angular momentum (OAM) of the LG modes. Tight confinements between a few hundred nanometers and several micrometers can be realized independently of the particle type. They are sufficient for performing many different spectroscopic measurements in situ in the trap (e.g. Raman scattering and microscopy, fluorescence, cavity ring-down spectroscopy and broadband light scattering). This opens new avenues to study the formation and ageing of atmospherically relevant absorbing aerosols such as BrC and SOA.

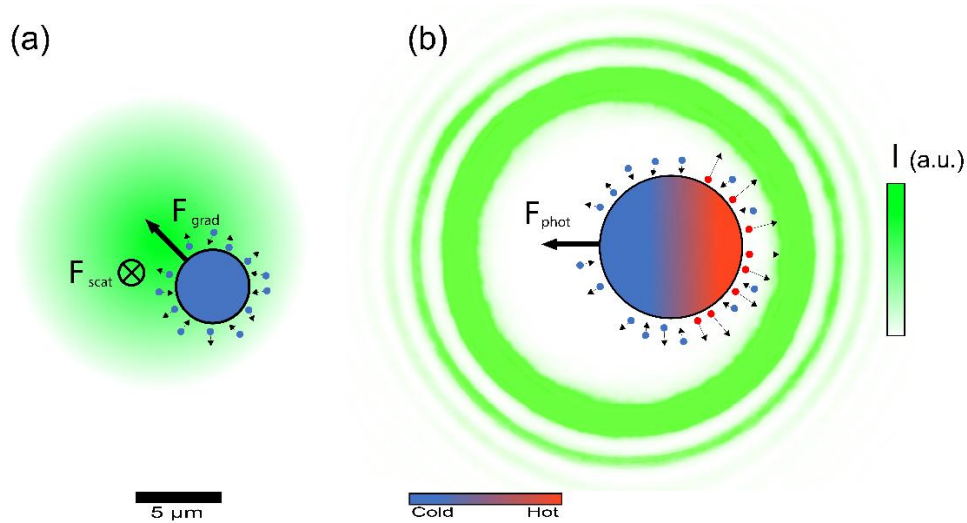

Fig. 1. Dominant trapping forces experienced by non-absorbing (panel (a)) and absorbing (panel (b)) particles. These are the gradient and scattering forces ($F_{grad}$ and $F_{sca}$) for non-absorbing particles and the photophoretic force $F_{phot}$ for absorbing particles. The intensity of the trapping laser beam (Gaussian beam in panel (a) and Laguerre-Gaussian beam in panel (b)) is shown by the gradient of the green color, while the temperature of the particle and surrounding gas molecules (small blue and red dots around the particle) is indicated by the blue to red color scale. In panel (b) the particle heats up inhomogeneously due absorption of the light when it approaches one side of the LG beam. The momentum transfer between the colliding gas molecules and the particle increases with increasing local temperature of the particle surface. This creates a PF opposite to the surface temperature gradient.

## 2. Methods

### 2.1 Universal trap

The trapping scheme is based on the use of LG laser modes with non-zero orbital angular momentum (OAM) to obtain vortex beams, which can trap absorbing particles. The trap is formed by four laser beams overlapping at the trapping position where they form two sets of counter-propagating beam pairs arranged perpendicular to each other (green beam in Fig. 2a). The overlap of the four laser beams with LG modes creates a light cage with the shape of a Steinmetz solid (Fig. 2b) which traps and confines the position of the particles. The size of the Steinmetz solid, and hence the confinement of the particle, is controlled by changing the OAM of the LG beams (Fig. 2c). The OAM of the LG beams is precisely controlled by the SLM. The mode of the laser beams can be dynamically tuned between Gaussian and LG of any OAM within each frame of the SLM (17 ms time resolution) without any optical or mechanical realignment.

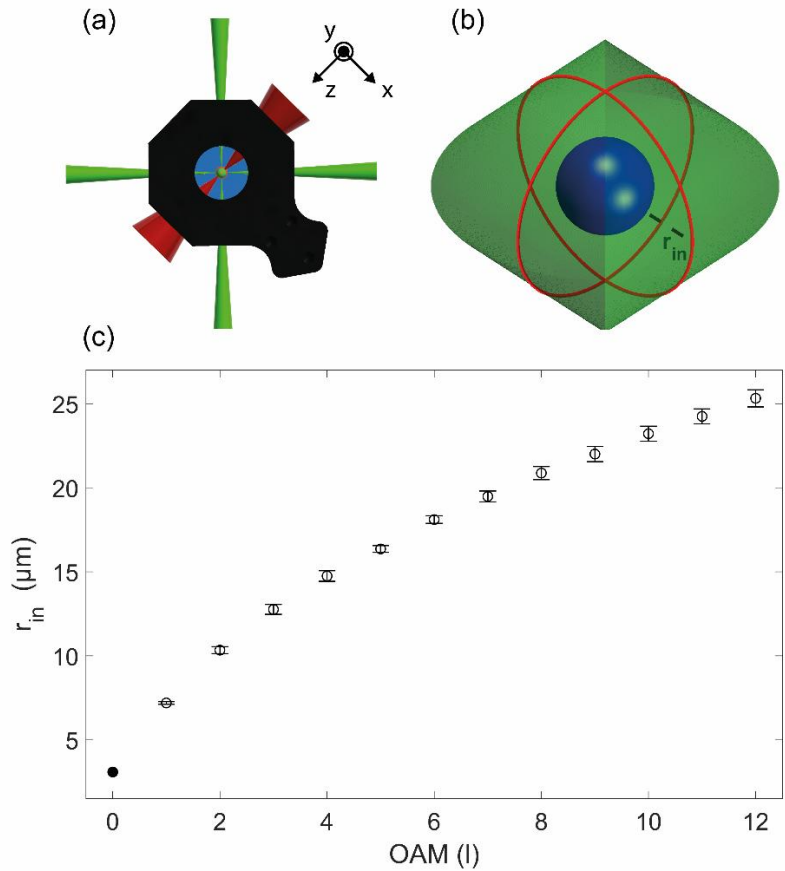

Fig. 2. Principle of the new universal trap for non-absorbing and absorbing particles. The four trapping beams (green beams in panel (a)) are arranged in two sets of counter-propagating beams, which are arranged perpendicularly to each other. The trapped particles are isolated in air inside a trapping cell (shown in black). The red beam represents the holographic imaging of the particle. The intersection of the four LG trapping beam creates a Steinmetz solid which shape is presented in panel (b). The inner radius of the Steinmetz solid ($r_{in}$) is the same along all directions parallel or perpendicular to the beams' propagation axes. Panel (c) shows the changes of $r_{in}$ as function of the topological charge $l$ of the OAM of the LG beams. The calculated $1/e^2$ diameter of the Gaussian beam (equivalent to LG beam with $l = 0$) is marked by a full black dot.

The experimental setup is presented in Fig. 3. A continuous wave (CW) Gaussian laser operating at 532 nm (Opus 532, Novanta Photonics) is expanded to 4.6 mm beam diameter using a two-lens telescope ($f_1$=40 mm, $f_1$=100 mm). The telescope contains a pinhole of 15 μm in the focal plane to clean the laser mode. The beam expansion is designed to fill out the LCOS display of the SLM (LUNA-VIS-111, HOLOEYE Photonics AG). To separate the phase-modulated beam from the unmodulated reflected beam, the 1st diffraction order of the fork grating imaged on the SLM is used in the subsequent trapping path. The phase modulated beam is again expanded by a factor of 2.5 using a telescope ($f_3$=50 mm, $f_4$=125 mm) to obtain a tighter focus of the laser beams at the trapping position. The beam is then divided into four arms with equal powers by three polarizing beamsplitters (PBS) in combination with three half-wave plates ($\lambda/2$). The four laser arms are focused into a custom-designed trapping cell (Fig. 2a) by the lenses $L_5$, $L_6$, $L_7$, and $L_8$ with focal lengths of 75 mm. To ensure that the four laser beams have the same diameter at the trapping position, all four trapping arms are built to have the same path length (i.e. the same convergence). Optical traps using four overlapping beams has been shown to provide higher confinement compared with single or two-beam traps[37, 57]. We tested the trapping of absorbing particles in a two- beam trap (one set of counter-propagating beams), but found unsatisfactory performance trapping stability and confinement for particles with different absorption strengths. When Gaussian mode beams are used, the trap is equivalent to a pair of two perpendicular counter-propagating tweezers, which are ideal for the trapping of non-absorbing particles relying on the usual scattering and gradient optical forces. When beams with LG modes ($|l|\geq 1$, where $l$ is the topological charge of the OAM) are used, absorbing particles are trapped by PF forces inside the Steinmetz solid formed by the overlapping hollow cores of the four vortex beams. The four-beam trap allows immobilization of both weakly and strongly absorbing particles (negative and positive PF, respectively) without any modification. The sign of the topological charge $l$ of the OAM, which changes with each reflection, needs to be opposite for each of the counter-propagating beams [58]. This conserves the OAM and ensures constructive interference in the trapping region. The output power of the trapping laser was set to ~110 mW. The power of each trapping beam in the trapping region was ~10 mW.

The particles were generated by a nebulizer (PARIBOY SX, PARI GmbH) and introduced into the trapping cell. A single droplet was trapped and investigated. Three types of aerosol droplets were studied: Droplets produced by nebulizing an aqueous solution of pure $K_2CO_3$ and aqueous solutions of $K_2CO_3$ with either fulvic acid (a BrC proxy) or nigrosin (an absorbing dye). Pure $K_2CO_3$ droplets are non-absorbing at the chosen trapping laser wavelength. The aqueous fulvic acid and nigrosin droplets (referred as fulvic acid/$K_2CO_3$ and nigrosin/$K_2CO_3$ droplets, respectively) are both absorbing, which cannot be trapped with Gaussian beams. The fulvic acid/$K_2CO_3$ stock solution consisted of 67% fulvic acid and 33% $K_2CO_3$ (solvent mass %). The nigrosin/$K_2CO_3$ stock solution, consisted of 33% nigrosin and 67% $K_2CO_3$. The amount of water of the stock solution was not relevant for the composition of the trapped droplets as they were equilibrated quickly by adjusting to the chosen surrounding relative humidity (RH), of ~50%. Despite the lower nigrosin content, nigrosin/$K_2CO_3$ droplets are more strongly absorbing compared with fulvic acid/$K_2CO_3$ droplets due to their larger imaginary part of the refractive index ($k$(fulvic acid/$K_2CO_3$~0.003 and $k$(nigrosin/$K_2CO_3$) ~0.06). The value of $k$ were determined from UV-VIS spectra (NanoDrop One, Thermo Fisher Scientific) of the nebulized stock solutions and extrapolation to the droplet concentration assuming equilibration to 50% RH using data from [29].

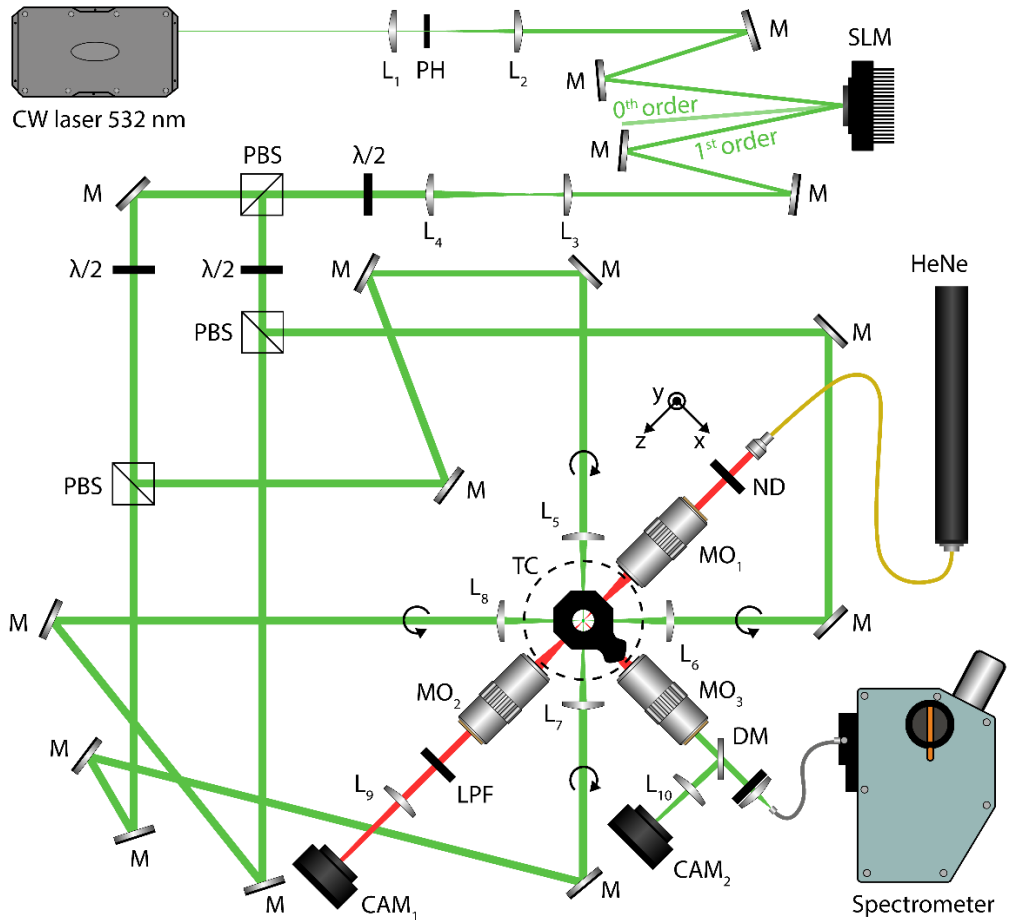

Fig. 3. Scheme of the experimental setup showing all lenses ($L_x$), mirrors (M), polarization beamsplitter cubes (PBS), half-wave plates ($\lambda/2$), microscope objectives ($MO_x$) and cameras ($CAM_x$), the spatial light modulator (SLM), pinhole (PH), long-pass filter (LPF), trapping cell (TC), neutral-density (ND) filter, dichroic mirror (DM) and HeNe laser for holography. The black arrows along each trapping beam indicates the direction of rotation of the LG beams.

For absorbing particles, Gaussian beams were used to provide an initial momentum transfer to the particles which pushed them into the trapping region. Gaussian beams were then changed to LG mode with comparatively high OAM (typically $|l| = 6$) to capture a single particle in the formed Steinmetz solid. The spatial confinement of the trap was then increased by smoothly decreasing the $l$ order of the beam. Aligning the 4-beam trap using LG beams with non-zero $l$ order would be challenging, but because no realignment is necessary between the different operation modes, the trap can be aligned using Gaussian beams and non-absorbing particles (aqueous $K_2CO_3$ droplets in our study). Making alignment straightforward.

### 2.2 Characterization techniques

The properties of the trapped droplets were characterized using digital in-line holography, Raman scattering / fluorescence spectroscopy (for fluorescing droplets) and broadband light scattering (BLS). The components for digital holography and Raman scattering are sketched in Fig. 3, while the BLS components are omitted to avoid over-crowding of the figure. More details can be found in previous publications of our group [X]. The digital in-line holography allows recording of the particle 3D position and quantifying its size and shape [59-62]. We used a HeNe laser (HNL050LB, Thorlabs and a microscope objective (M Plan APO 20X, Mitutoyo, numerical aperture (NA) = 0.42) to focus the beam in front of the trapped particle. A second microscope objective of the same type is used to collect the hologram created by the interference of the particle scattering and the reference incident laser. A lens $L_9$ ($f_9$=75 mm) is used to fill out the sensor (800×600 pixels) of the CMOS camera 1 (ace acA800-510uc, Basler AG and Beam Profiler 4M, Edmund Optics). The spatial resolution of the holographic imaging is ~0.9 μm. To avoid possible interference coming from the scattering of the trapping beams by the particle, a 532 nm notch filter is placed in front of the holography camera. The numerical reconstruction of in-line holograms was done using a two Fourier transform approach [59]. The confinement study of salt was done with 852 holograms. For the probability distribution maps of fulvic acid/$K_2CO_3$ 9002 holograms were taken for each LG mode. The nigrosin/$K_2CO_3$ confinement study was done with 480 and 900 holograms for $|l| = 5$, $|l| = 6$ respectively.

For Raman scattering and fluorescence spectroscopy, the trapping laser was used as excitation source. The elastic and inelastic light emitted by the trapped particle were collected with a microscope objective (M Plan APO 20X, Mitutoyo, NA=0.42). A dichroic mirror and two 532 nm notch filters were used to detect only the inelastic light (Raman scattering and fluorescence, if there is any). The scattered/fluoresced light was detected with a spectrometer (Shamrock SR303iA, Oxford Instruments Andor) and coupled to a high sensitivity camera (Newton DU970p-UVB). More can be found in [29].

The BLS measurements provide accurate information on the droplet size for non-absorbing (here pure aqueous $K_2CO_3$ droplets) following the procedure described in [38]. To excite BLS, we used the light of a broadband Xenon lamp (HPX-2000, Ocean Insight), focused on the droplet with a microscope objective (M Plan APO 20X, Mitutoyo, NA=0.42). The elastic light scattered by the droplet in the wavelength range between 370 and 700 nm was collected with another objective (M Plan APO 20X, Mitutoyo, NA=0.42) and detected with a spectrometer (Maya 2000-PRO, Ocean Optics). The intensity pattern of the scattered light was fitted with Mie theory to retrieve the droplet radius and wavelength-dependent refractive index [38].

## 3. Results

3.1 Particle confinement in the Gaussian and LG beam trap

The spatial confinement measured with digital holography are summarized in Fig. 4 for non-absorbing pure $K_2CO_3$ droplets inside the Gaussian beam trap and in Fig. 5 for the different absorbing droplets in the LG beam trap. All data are plotted as probability distribution maps representing the position of the center of the droplet in the holography reconstruction plan during the motion of the droplet in the trap. The confinement along the propagation direction of the holography beam (Z-axis shown in Fig. 2 and 3) is similar to the confinement measured along the X-axis because of the symmetry of the light in trapping region (see Fig. 2 for the example of the symmetry Steinmetz solid). This holds for both Gaussian and LG beams. The measured probability distributions of the droplet position are governed from the combined effects of the steepness of the trapping potential, which tries to keep the particle at the equilibrium position in the center of the potential, and the Brownian force exerted by the momentum transfer from the surrounding molecules to the trapped droplet. Without a trap, the Brownian force creates random Brownian motion of the droplet. In a trap, Brownian motion becomes spatially confined. Steeper trapping potentials lead to a better spatial confinement due to more efficient counter-acting of the Brownian motion. Figure 4a shows the probability distribution corresponding to 24 different non-absorbing aqueous $K_2CO_3$ droplets for Gaussian trapping beams. A Gaussian fit of the probability distribution of the particle position were performed along X- and Y-axis to quantify the confinement along these axes. The standard deviation of the Gaussian fit along X- and Y-axis ($\sigma_x$ and $\sigma_y$ respectively) are used as the measure of the particle confinement. Inside the Gaussian beam trap, we determined values of $\sigma_x$ and $\sigma_y$ of 0.7 and 0.6 µm, respectively. Figure 4b shows the corresponding size distribution of the 24 measured aqueous droplets. No significant dependence of the confinement on the droplet size was observed for droplet radii between 0.5 and 2.5 µm.

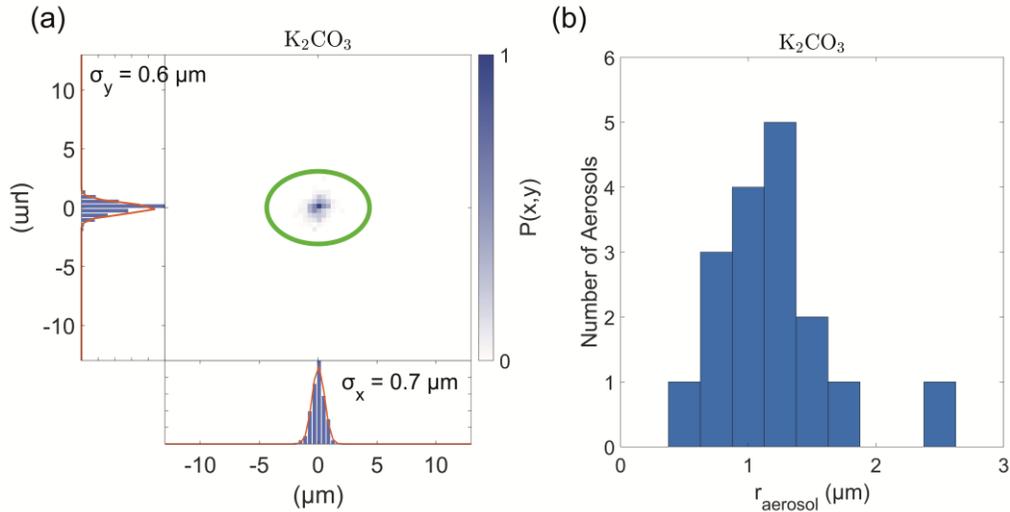

Fig. 4. Panel (a): Probability distribution maps of the center position of the motion of single non-absorbing $K_2CO_3$ droplets in a Gaussian beam trap. The green ellipse indicates the size of the Gaussian beam cut along reconstruction plane of the holography ($1/e^2$ diameter of the Gaussian beams is equal to 6.17 µm at their focal plane). The ellipsoidal shape of the light beam in the reconstruction of the holography is due to the 45° angle between the propagation axis of the trapping beams and the holography axis. Gaussian fits along the x and y axis quantify the trapping confinement of the droplets. The obtained standard deviation ($\sigma_x$ and $\sigma_y$) are indicated in the graph. They quantify the confinement. Panel (b) shows the size distribution of the 24 single droplets from which the probability distribution in panel (a) was determined.

Figure 5(a-d) shows the results of the trapping confinement study for absorbing aqueous fulvic acid/$K_2CO_3$ droplets. They were trapped in vortex LG beams with different $l$ orders. The results show that the confinement increases with decreasing $|l|$. $\sigma_x$ decreases from 6.1 µm for $|l| = 5$ to $\sigma_x$ 0.7 µm for $|l| = 1$. Figures 5(e) and 5(f) quantify the trapping confinement of strongly absorbing aqueous nigrosin droplets for $|l| = 5$ and 6, respectively. The measurements in Fig. 5 illustrate that the confinement can be controlled by changing $|l|$ of the LG beams. The comparison of Fig. 5(d) and 5(e) shows that particles with higher light absorption (nigrosin/$K_2CO_3$ droplets are more absorbing than the fulvic acid/$K_2CO_3$ droplets) are more confined in the trap at a given $|l|$, even though their size is significantly smaller (smaller size lead to a better confinement when all other parameters are the same). The confinement of absorbing droplets can be increased by decreasing the azimuthal order $l$. The caveat of the increased confinement is the higher exposure to light of the particles, which is not an issue if the compounds of the trapped particle are more or less photostable and thermally stable. For thermally instable droplets, light absorption may result in evaporate the droplet. For droplets containing compounds that are not photostable, the increase of light exposure due to the decrease of $|l|$ will accelerate the photochemical reaction in the trapped droplet.

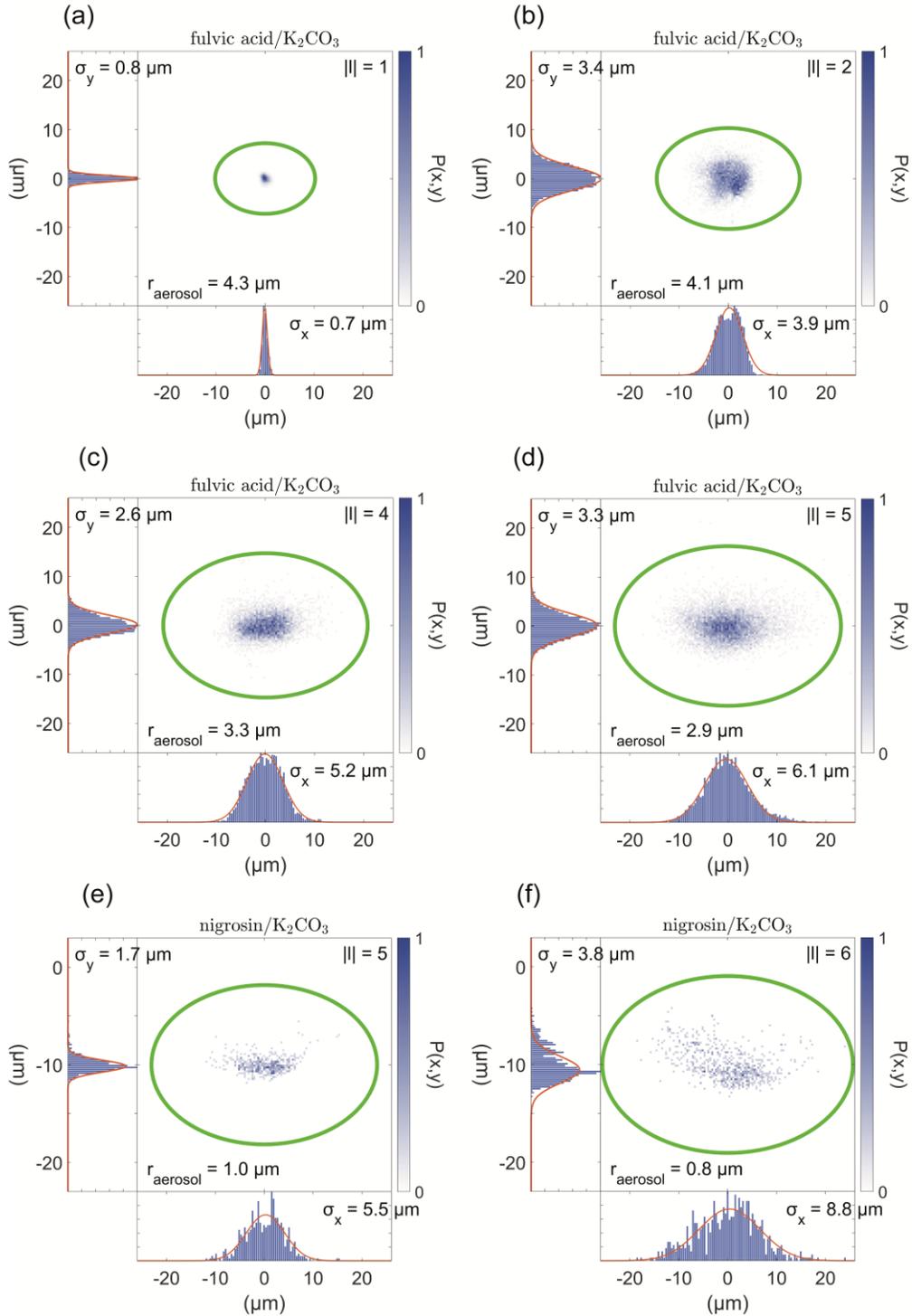

Fig. 5. Probability distribution maps of the position of the motion of single absorbing droplets in LG beam traps with different |l| orders. Gaussian fits along the x and y axis quantify the trapping confinement of the droplets. The radius of the droplets (determined by holography) is indicated in the bottom left of each panel. Panels (a-d) presents the results for absorbing

aqueous fulvic acid/$K_2CO_3$ droplets for $|l|$ =1, 2, 4 and 5, respectively Panels (f) and (e) presents the results for highly absorbing aqueous nigrosin/$K_2CO_3$ droplets for $|l|$ =5 and 6, respectively. The green ellipses indicate the size of the LG beams cut along the reconstruction plane of the digital holography imaging. The ellipsoidal shape of the light beam in the holography reconstruction plane is due to the 45° angle between the propagation axis of the trapping beams and the holography axis.

3.2 Raman spectroscopy

Figure 6(a-c) shows Raman and fluorescence spectra of pure aqueous $K_2CO_3$ droplets (trapped with Gaussian beam), aqueous fulvic acid/$K_2CO_3$ droplets and aqueous nigrosin/$K_2CO_3$ droplets (both trapped with LG beams), respectively. Figures 6(b) and (c) reveal Raman and fluorescence spectra can be recorded even in LG beams with reasonable time resolution (typical integration of time of 2 min). The spectra of the pure $K_2CO_3$ droplets and the nigrosin/$K_2CO_3$ droplets are look similar and show no time dependence. In all spectra, the peaks around 2330 cm$^{-1}$ and 3000 cm$^{-1}$ correspond to the Raman scattering of the gaseous $N_2$ and the elastic scattering of the holography beam, respectively. The peak observed at 2430 cm$^{-1}$ is an artifact due to spectrometer which is visible even without trapped droplet. The other very sharp peaks are due to cosmic rays. The broad band around 3300 cm$^{-1}$ arises from the O-H stretching vibration of water. The peak at 1065 cm$^{-1}$ is due to the stretching vibration of the $CO_3^{2-}$ ion. The spectrum of the fulvic acid/$K_2CO_3$ droplet is dominated by fluorescence coming from the fulvic acid molecules [63] (Fig. 6b), which creates the very broad background (from 500 to 3500 cm$^{-1}$). Strong bleaching of the fluorescence and hence decomposition of fulvic acid molecules is observed as a function of time. Bleaching results in the disappearance of the broad background. For the 4 fulvic acid droplets studied here, the fluorescence decreased on average by a factor of $10 \pm 2$ (standard deviation of the decay) within 5 minutes. The complete bleaching also shows that droplets can be trapped during the photochemical reaction until all the fulvic acid molecules have reacted away and the droplets have become non-absorbing, demonstrating the suitability of our new universal trap for studying photochemical reaction in particles.

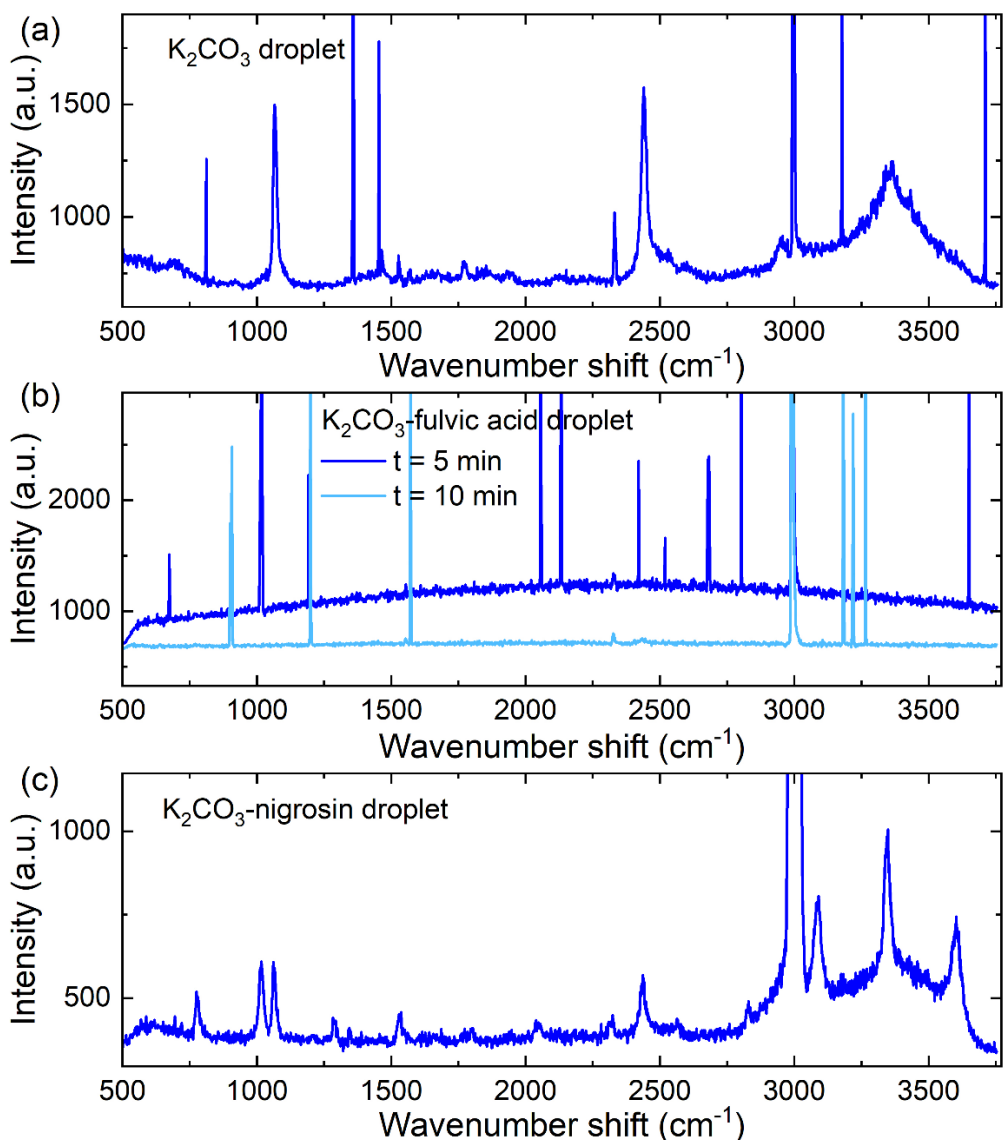

Fig. 6. Typical Raman and/or fluorescence spectra of pure aqueous $K_2CO_3$ droplets (panel (a)), aqueous fulvic acid/$K_2CO_3$ droplets (panel (b)) and an aqueous nigrosin/ $K_2CO_3$ droplets (panel (c)). The time evolution of the spectra is only shown in panel (b) because the spectra of the other droplets where time independent.

## 4. Conclusion

Our new universal optical trap allows trapping of particles with any light absorption strength and provides tight confinement (between a few hundred nanometers and a few micrometers) for particles with sizes in the submicrometer to micrometer range - without the need of any mechanical realignment. It uses fundamental Gaussian beams to trap non-absorbing particles and vortex LG beams to trap weakly to highly absorbing particles. Tuning between the two operation modes by changing the mode of the laser with a SLM takes less than 20 ms. The trap also allows to study strongly absorbing droplets, which to the best of our knowledge has not been demonstrated before. This opens new avenues for single particle studies, such as the investigations of aging of single SOA and BrC aerosols. The presented work and results demonstrate these new capabilities. Non-absorbing pure aqueous $K_2CO_3$ droplets were trapped as well as absorbing aqueous droplets containing fulvic acid, a common BrC compound, and nigrosin, an absorbing dye. Fulvic acid droplets could be trapped throughout the photochemical degradation of fulvic acid, i.e. when the droplets change from absorbing to non-absorbing. The confinement of the trapped particles can be controlled by changing the |*l*| orders of the LG trapping beams. For |*l*| =1, the particles are confined within approximately 1 μm of the equilibrium trapping position, while for |*l*| =5, the particles are confined within approximately 6 μm. These tight confinements allow to perform in-trap spectroscopic and microscopic studies for particle characterization.

Acknowledgement
We thank P. Albrecht and M. Steger for technical support. We acknowledge funding by the Swiss National Science Foundation (grant nr. 200021-236446).

**References**

1. "Technical Summary," in *Climate Change 2021 – The Physical Science Basis: Working Group I Contribution to the Sixth Assessment Report of the Intergovernmental Panel on Climate Change*, C. Intergovernmental Panel on Climate, ed. (Cambridge University Press, 2023), pp. 35-144.
2. R. Saleh, "From Measurements to Models: Toward Accurate Representation of Brown Carbon in Climate Calculations," Curr. Pollut. Rep. **6**, 90-104 (2020).
3. F. Mayyas, H. Aldawod, K. H. Alzoubi et al., "Comparison of the cardiac effects of electronic cigarette aerosol exposure with waterpipe and combustible cigarette smoke exposure in rats," Life Sciences **251**, 117644 (2020).
4. G. David, E. A. Parmentier, I. Taurino et al., "Assessment of the Chemical Evolution of E-Cigarette Droplets," CHIMIA International Journal for Chemistry **74**, 733-733 (2020).
5. U. Pöschl, and M. Shiraiwa, "Multiphase Chemistry at the Atmosphere–Biosphere Interface Influencing Climate and Public Health in the Anthropocene," Chem. Rev. **115**, 4440-4475 (2015).
6. T. Arfin, A. M. Pillai, N. Mathew et al., "An overview of atmospheric aerosol and their effects on human health," Environmental Science and Pollution Research **30**, 125347-125369 (2023).
7. J. F. Kok, T. Storelvmo, V. A. Karydis et al., "Mineral dust aerosol impacts on global climate and climate change," Nature Reviews Earth & Environment **4**, 71-86 (2023).
8. A. Laskin, C. P. West, and A. P. S. Hettiyadura, "Molecular insights into the composition, sources, and aging of atmospheric brown carbon," Chem. Soc. Rev. **54**, 1583-1612 (2025).
9. J. H. Vincent, *Aerosol Science for Industrial Hygienists* (Pergamon, 1995).
10. M. R. Marvin, P. I. Palmer, F. Yao et al., "Uncertainties from biomass burning aerosols in air quality models obscure public health impacts in Southeast Asia," Atmos. Chem. Phys. **24**, 3699-3715 (2024).
11. M. Kahnert, T. Nousiainen, and H. Lindqvist, "Models for integrated and differential scattering optical properties of encapsulated light absorbing carbon aggregates," Opt. Express **21**, 7974-7993 (2013).
12. X. Cao, J. Liu, Y. Wu et al., "A Review on Brown Carbon Aerosol in China: From Molecular Composition to Climate Impact," Curr. Pollut. Rep. **10**, 326-343 (2024).
13. R. F. Hems, E. G. Schnitzler, C. Liu-Kang et al., "Aging of Atmospheric Brown Carbon Aerosol," ACS Earth Space Chem. **5**, 722-748 (2021).
14. Y. Zhang, Q. Zhang, N. Wu et al., "Weakened Haze Mitigation Induced by Enhanced Aging of Black Carbon in China," Environ. Sci. Technol. **56**, 7629-7636 (2022).
15. K. J. Zarzana, D. O. De Haan, M. A. Freedman et al., "Optical Properties of the Products of α-Dicarbonyl and Amine Reactions in Simulated Cloud Droplets," Environ. Sci. Technol. **46**, 4845-4851 (2012).
16. D. O. De Haan, L. N. Hawkins, H. G. Welsh et al., "Brown Carbon Production in Ammonium- or Amine-Containing Aerosol Particles by Reactive Uptake of Methylglyoxal and Photolytic Cloud Cycling," Environ. Sci. Technol. **51**, 7458-7466 (2017).
17. A. K. Y. Lee, R. Zhao, R. Li et al., "Formation of Light Absorbing Organo-Nitrogen Species from Evaporation of Droplets Containing Glyoxal and Ammonium Sulfate," Environ. Sci. Technol. **47**, 12819-12826 (2013).
18. A. Laskin, J. Laskin, and S. A. Nizkorodov, "Chemistry of Atmospheric Brown Carbon," Chem. Rev. **115**, 4335-4382 (2015).
19. N. Paisi, J. Kushta, G. Georgiou et al., "Modeling of carbonaceous aerosols for air pollution health impact studies in Europe," Air Quality, Atmosphere & Health **17**, 2091-2104 (2024).
20. H. O. T. Pye, C. K. Ward-Caviness, B. N. Murphy et al., "Secondary organic aerosol association with cardiorespiratory disease mortality in the United States," Nat. Commun. **12**, 7215 (2021).

21. K. S. Hopstock, Q. Xie, M. A. Alvarado et al., "Molecular Characterization and Photoreactivity of Organic Aerosols Formed from Pyrolysis of Urban Materials during Fires at the Wildland–Urban Interface," ACS ES&T Air **1**, 1495-1506 (2024).
22. J. M. Anglada, M. T. C. Martins-Costa, J. S. Francisco et al., "Photoinduced Oxidation Reactions at the Air–Water Interface," J. Am. Chem. Soc. **142**, 16140-16155 (2020).
23. A. Athanasiadis, C. Fitzgerald, N. M. Davidson et al., "Dynamic viscosity mapping of the oxidation of squalene aerosol particles," Phys. Chem. Chem. Phys. **18**, 30385-30393 (2016).
24. J. Zhong, C. Zhu, L. Li et al., "Interaction of SO2 with the Surface of a Water Nanodroplet," J. Am. Chem. Soc. **139**, 17168-17174 (2017).
25. S. Banerjee, E. Gnanamani, X. Yan et al., "Can all bulk-phase reactions be accelerated in microdroplets?," Analyst **142**, 1399-1402 (2017).
26. K. J. Kappes, A. M. Deal, M. F. Jespersen et al., "Chemistry and Photochemistry of Pyruvic Acid at the Air–Water Interface," J. Phys. Chem. A **125**, 1036-1049 (2021).
27. E. Antonsson, C. Raschpichler, B. Langer et al., "Surface Composition of Free Mixed NaCl/Na2SO4 Nanoscale Aerosols Probed by X-ray Photoelectron Spectroscopy," J. Phys. Chem. A **122**, 2695-2702 (2018).
28. M. Nirmal, B. O. Dabbousi, M. G. Bawendi et al., "Fluorescence intermittency in single cadmium selenide nanocrystals," Nature **383**, 802-804 (1996).
29. K. Esat, G. David, T. Poulkas et al., "Phase transition dynamics of single optically trapped aqueous potassium carbonate particles," Phys. Chem. Chem. Phys. **20**, 11598-11607 (2018).
30. M. A. B. Andrade, N. Pérez, and J. C. Adamowski, "Review of Progress in Acoustic Levitation," Brazilian Journal of Physics **48**, 190-213 (2018).
31. E. J. Davis, and A. K. Ray, "Single aerosol particle size and mass measurements using an electrodynamic balance," J. Colloid Interface Sci. **75**, 566-576 (1980).
32. A. K. Y. Lee, and C. K. Chan, "Single particle Raman spectroscopy for investigating atmospheric heterogeneous reactions of organic aerosols," Atm. Env. **41**, 4611-4621 (2007).
33. I. N. Tang, A. C. Tridico, and K. H. Fung, "Thermodynamic and optical properties of sea salt aerosols," J. Geophys. Res. **102**, 23269-23275 (1997).
34. A. Ashkin, "Acceleration and Trapping of Particles by Radiation Pressure," Phys. Rev. Lett. **24**, 156-159 (1970).
35. M. D. Summers, J. P. Reid, and D. McGloin, "Optical guiding of aerosol droplets," Opt. Express **14**, 6373-6380 (2006).
36. C. Wang, Y.-L. Pan, and G. Videen, "Optical trapping and laser-spectroscopy measurements of single particles in air: a review," Measurement Science and Technology **32**, 102005 (2021).
37. G. David, K. Esat, S. Hartweg et al., "Stability of aerosol droplets in Bessel beam optical traps under constant and pulsed external forces," J. Chem. Phys. **142**, 154506 (2015).
38. G. David, K. Esat, I. Ritsch et al., "Ultraviolet broadband light scattering for optically-trapped submicron-sized aerosol particles," Phys. Chem. Chem. Phys. **18**, 5477-5485 (2016).
39. S. H. Jones, M. D. King, and A. D. Ward, "Atmospherically relevant core-shell aerosol studied using optical trapping and Mie scattering," Chem. Commun. **51**, 4914-4917 (2015).
40. T. Li, S. Kheifets, D. Medellin et al., "Measurement of the Instantaneous Velocity of a Brownian Particle," Science **328**, 1673-1675 (2010).
41. Y. Yang, Y. Ren, M. Chen et al., "Optical trapping with structured light: a review," Advanced Photonics **3**, 034001 (2021).
42. Z. Gong, Y.-L. Pan, G. Videen et al., "Optical trapping and manipulation of single particles in air: Principles, technical details, and applications," J. Quant. Spectrosc. Radiat. Transfer **214**, 94-119 (2018).
43. H. Rubinsztein-Dunlop, T. A. Nieminen, M. E. J. Friese et al., "Optical Trapping of Absorbing Particles," in *Advances in Quantum Chemistry*, P.-O. Löwdin, ed. (Academic Press, 1998), pp. 469-492.
44. H. He, M. E. J. Friese, N. R. Heckenberg et al., "Direct Observation of Transfer of Angular Momentum to Absorptive Particles from a Laser Beam with a Phase Singularity," Phys. Rev. Lett. **75**, 826-829 (1995).
45. A. Ashkin, "Forces of a single-beam gradient laser trap on a dielectric sphere in the ray optics regime," Biophysical Journal **61**, 569-582 (1992).
46. S. H. Jones, M. D. King, and A. D. Ward, "Determining the unique refractive index properties of solid polystyrene aerosol using broadband Mie scattering from optically trapped beads," Phys. Chem. Chem. Phys. **15**, 20735-20741 (2013).
47. M. I. Cotterell, T. C. Preston, A. J. Orr-Ewing et al., "Assessing the accuracy of complex refractive index retrievals from single aerosol particle cavity ring-down spectroscopy," Aerosol Sci. Technol. **50**, 1077-1095 (2016).
48. L. Rkiouak, M. J. Tang, J. C. J. Camp et al., "Optical trapping and Raman spectroscopy of solid particles," Phys. Chem. Chem. Phys. **16**, 11426-11434 (2014).
49. O. Jovanovic, "Photophoresis—Light induced motion of particles suspended in gas," J. Quant. Spectrosc. Radiat. Transfer **110**, 889-901 (2009).
50. V. G. Shvedov, A. S. Desyatnikov, A. V. Rode et al., "Optical vortex beams for trapping and transport of particles in air," Applied Physics A **100**, 327-331 (2010).
51. Z. Yang, N. T. Tsona, C. George et al., "Nitrogen-Containing Compounds Enhance Light Absorption of Aromatic-Derived Brown Carbon," Environ. Sci. Technol. **56**, 4005-4016 (2022).

52. Y. Zhou, C. P. West, A. P. S. Hettiyadura et al., "Molecular Characterization of Water-Soluble Brown Carbon Chromophores in Snowpack from Northern Xinjiang, China," Environ. Sci. Technol. **56**, 4173-4186 (2022).
53. X. Hu, Z. Guo, W. Sun et al., "Atmospheric Processing of Particulate Imidazole Compounds Driven by Photochemistry," Environmental Science & Technology Letters **9**, 265-271 (2022).
54. P. Lin, N. Bluvshtein, Y. Rudich et al., "Molecular Chemistry of Atmospheric Brown Carbon Inferred from a Nationwide Biomass Burning Event," Environ. Sci. Technol. **51**, 11561-11570 (2017).
55. X. Lian, G. Zhang, Y. Yang et al., "Evidence for the Formation of Imidazole from Carbonyls and Reduced Nitrogen Species at the Individual Particle Level in the Ambient Atmosphere," Environmental Science & Technology Letters **8**, 9-15 (2021).
56. Y. Liu, T. Wang, X. Fang et al., "Brown carbon: An underlying driving force for rapid atmospheric sulfate formation and haze event," Science of The Total Environment **734**, 139415 (2020).
57. I. Thanopulos, D. Luckhaus, T. C. Preston et al., "Dynamics of submicron aerosol droplets in a robust optical trap formed by multiple Bessel beams," J. Appl. Phys. **115**, 154304 (2014).
58. V. G. Shvedov, A. S. Desyatnikov, A. V. Rode et al., "Optical guiding of absorbing nanoclusters in air," Opt. Express **17**, 5743-5757 (2009).
59. T. Latychevskaia, and H.-W. Fink, "Practical algorithms for simulation and reconstruction of digital in-line holograms," Appl. Opt. **54**, 2424-2434 (2015).
60. M. J. Berg, and G. Videen, "Digital holographic imaging of aerosol particles in flight," J. Quant. Spectrosc. Radiat. Transfer **112**, 1776-1783 (2011).
61. S. K. Jericho, J. Garcia-Sucerquia, W. Xu et al., "Submersible digital in-line holographic microscope," Rev. Sci. Instrum. **77**, 043706 (2006).
62. G. David, K. Esat, I. Thanopulos et al., "Digital holography of optically-trapped aerosol particles," Commun. Chem. **1**, 46 (2018).
63. C. Coelho, G. Guyot, A. ter Halle et al., "Photoreactivity of humic substances: relationship between fluorescence and singlet oxygen production," Environmental Chemistry Letters **9**, 447-451 (2011).